\documentclass[10pt,conference]{IEEEtran}

\usepackage[utf8]{inputenc}
\usepackage[T1]{fontenc} %optional
\usepackage{amsmath}
\usepackage{graphicx}
\usepackage[cmintegrals]{newtxmath}
\usepackage{bm} % optional
\usepackage{cite}
\usepackage{booktabs}
\usepackage{array}
\usepackage{url}
\usepackage[ruled]{algorithm2e}
\usepackage{multirow}
\newcolumntype{L}[1]{>{\raggedright\let\newline\\\arraybackslash\hspace{0pt}}m{#1}}
\newcommand{\pluseq}{\mathrel{+}=}
\DeclareMathOperator*{\argmin}{arg\,min}

\title{RANK: AI-assisted End-to-End Architecture for Detecting Persistent Attacks in Enterprise Networks}
\author{\IEEEauthorblockN{Hazem M. Soliman, Geoff Salmon, Du{\v s}an Sovilj, Mohan Rao}
\IEEEauthorblockA{\textit{Arctic Wolf Networks} \\
Waterloo, Canada \\
\{hazem.soliman,geoff.salmon,dusan.sovilj,mohan.rao\}@arcticwolf.com}}
\date{January 2020}

\begin{document}

\maketitle

\begin{abstract}
Advanced Persistent Threats (APTs) are sophisticated multi-step attacks, planned and executed by skilled adversaries targeting modern government and enterprise networks. Intrusion Detection Systems (IDSs) and User and Entity Behavior Analytics (UEBA) are commonly employed to aid a security analyst in the detection of APTs. The prolonged nature of APTs, combined with the granular focus of UEBA and IDS, results in overwhelming the analyst with an increasingly impractical number of alerts. Consequent to this abundance of data, and together with the crucial importance of the problem as well as the high cost of the skilled personnel involved, the problem of APT detection becomes a perfect candidate for automation through Artificial Intelligence (AI). In this paper, we provide, up to our knowledge, the first study and implementation of an end-to-end AI-assisted architecture for detecting APTs -- RANK. The goal of the system is not to replace the analyst, rather, it is to automate the complete pipeline from data sources to a final set of incidents for analyst review. The architecture is composed of four consecutive steps: 1) alert templating and merging, 2) alert graph construction, 3) alert graph partitioning into incidents, and 4) incident scoring and ordering. We evaluate our architecture against the 2000 DARPA Intrusion Detection dataset, as well as a read-world private dataset from a medium-scale enterprise. Extensive results are provided showing a three order of magnitude reduction in the amount of data to be reviewed by the analyst, innovative extraction of incidents and security-wise scoring of extracted incidents.
\end{abstract}


\section{Introduction}

Advanced Persistent Threats (APTs) form a crucial part of the attack campaigns targeting enterprise networks \cite{Ghafir2014}. A combination of stealth actions, sophisticated exploits and long-term operation marks them as one of the most dangerous and hardest to detect cybersecurity attacks \cite{mcwhorter2013apt1}. As a planned multi-step attack, APTs do not target a single vulnerability. Instead, an APT typically starts by gaining access to the network, exploiting several vulnerabilities, and remaining silent for extended periods of time \cite{alvarez2017ibm}.  With data exfiltration as the common end-goal of these attacks, the economic damage is particularly severe due to the risk of losing intellectual properties or sensitive data \cite{berghel2017equifax}. Various companies and government agencies have fallen victim to these attacks \cite{rivner2011anatomy, caplan2013cyber}. In light of their severity and impact, detecting APTs has increasingly become the focus of both academia and industry.

A pattern is emerging in the literature where the detection of APTs is composed of numerous small and simple detectors each focused on a single step in the APT chain. Examples of this methodology are \cite{sexton2015attack}, \cite{cao2019preempting}. These small detectors are based on either signature-based intrusion detection systems (IDS), user entity behavior analytics (UEBA) \cite{shashanka2016user}, or a combination of both. However, IDS and UEBA have well-documented problems such as high false-positive ratio, expensive cost of error, difficulty of explainability for anomaly-based detection, lack of training data and continuously improving intelligent adversaries \cite{sommer2010outside}. APT-based detection methods attempt to solve some of those issues in order to significantly reduce the false-positive ratio. The main premise behind these methods is that an alert is malicious if it is part of a larger group of correlated alerts that together represent a proper APT plan.

A leading effort to study APT attack plans is the ATT$\&$CK matrix developed by the MITRE corporation \cite{strom2018mitre}. The study observed that attacks tend to follow similar patterns, and provided a taxonomy describing both offensive and defensive operations. The matrix describes the tactics, commonly known as the attack phases, the techniques for carrying out a tactic, and the observed instances of such techniques in known attacks.  Detection and Extraction of ATT\&CK-aware security incidents has been one of our major design goals.
% Our main contribution in this paper is developing an end-to-end architecture for detecting ATT\&CK-aware security incidents.

In this paper, we propose RANK, an end-to-end architecture for detecting persistent attacks by incorporating numerous alerts in a meaningful and automated approach. The architecture is composed of several steps:

\begin{enumerate}
  \item \textbf{Alert templating and merging}: In this step, the aim is to reduce the number of generated alerts by merging together the ones pertaining to the same step of the attack. The new set of merged alerts is referred to as generalized alerts.
  \item \textbf{Alert graph construction}: Once the number of alerts is reduced, we establish relationships between remaining alerts according to their MITRE tactics, and form an alert graph representing those relationships.
  \item \textbf{Alert graph partitioning}: The partitioning part focuses on finding the smallest subgraphs within the alert graph that correspond to concise incidents. We refer to these as incident graphs. 
  \item \textbf{Incident scoring}: Scores are assigned to each incident graph found in previous step, utilizing factor graphs (FG) and alert scores. For each incident graph, we can compute the likelihood of presence for all MITRE tactics, a useful and important summary. Both the incident graph and the summary are finally delivered to the security analyst. 
\end{enumerate}

% Moreover, we also discuss how the MITRE tactics can be incorporated into the alert graph construction and partitioning, as well as the scoring step. 
The proposed architecture is also able to integrate analyst feedback as part of incident investigation and threat hunting. This is enabled by the evidence-based queries provided by the scoring factor graph for each incident. 

To summarize, our contributions in this paper are:
\begin{itemize}
    \item We propose the first, up to our knowledge, architecture for end-to-end detection of persistent attacks from network and endpoint data.
    \item We discuss a novel approach for partitioning an alert graph into separate incidents and how it fits into the architecture, which is an extension of our previous work \cite{Soliman2020}.
    \item We detail various practical aspects of the involved algorithms, such as meaningful clustering of security alerts, calculating correlation between IP addresses, and building all the algorithms with minimal manual tweaking effort due to the lack of labelled datasets.
    \item The proposed architecture significantly reduces the investigation burden on the security analyst. For example, our experiments show that more than 70 thousand alerts can be distilled into only 60 incidents. This is a considerably more concise, yet equally informative, summary of the data.
\end{itemize}

The remainder of the paper is organized as follows: in Section \ref{sec:rel_work} we provide a summary of related works in the literature. Section \ref{sec:model} provides an overview of the proposed architecture and its components. These components are then detailed in the following sections: \ref{sec:template}, \ref{sec:corrgraph}, \ref{sec:graphpartition} and \ref{sec:scoring}. Experimental results are provided in Section \ref{sec:results}. A discussion of the related peripheral problems is provided in Section \ref{sec:periph}. Section \ref{sec:future} discusses potential future work, and the paper is concluded in Section \ref{sec:conclusion}.

\section{Related work}
\label{sec:rel_work}

In the attempt to detect and reconstruct sophisticated multi-step APTs, individual intrusion alerts are not deemed as a sufficient indication of compromise \cite{cao2015preemptive}. Instead of individual alerts, more advanced detection approaches try to reconstruct the attack plan \cite{qin2005probabilistic}. In the literature, these approaches can be divided into two broad categories: the first is based on Fisher's method for $p\text{-value}$ combination \cite{sexton2015attack}; while the second is based on the concept of attack graphs \cite{eshete2016attack}. This first approach of detecting APTs assigns a joint score for multiple detections, and has the advantage of simplicity. However it lacks the sequencing capabilities of attack graphs. In other words, the first approach only considers the coverage of MITRE tactics in the detections, and does not consider whether the order of said detections actually makes sense as an attack plan. Attack graphs are a more powerful representation of APTs \cite{eshete2016attack}. Besides their visually attractive properties, the topology of a directed graph is a representation of the attack plan progression. In such graphs, each vertex represents an individual alert, which might come from a variety of security solutions, and each edge represents the progression between two alerts \cite{cao2019preempting}. With their vertices comprised of security alerts, attack graphs are also sometimes called alert graphs, which is the terminology we use in this work.

Despite its advantageous features, an attack graph containing all alerts is usually overwhelmingly large due to the high false-positive rate of the individual alert detectors. This problem can be addressed in two ways. The first approach combines alerts corresponding to the same event together into a Hyper/Generalized alert \cite{julisch2003clustering}, while the second approach works by partitioning the alert graph into smaller sub-graphs. Both approaches aim at producing smaller alert graphs in order to reduce the investigation burden on the security analyst. An example of the second approach is given in \cite{haas2019alert}, where a community-detection approach is proposed with each resulting community corresponding to a single attack step. In \cite{ning2002constructing}, alert prerequisites and consequences are defined for each alert to limit the number of edges. However, this approach requires heavy manual work, and can not handle cases when multiple attacks are carried-out simultaneously. Finally, in \cite{cao2019preempting} a separate graph is defined for each single asset. This approach results in manageable graphs, but it ignores lateral movement between assets, a common tactic in APT attacks. 

It is worth emphasizing again the difference between the two main approaches in reducing the size of the graph. The first approach tries to combine together alerts corresponding to the same step in an APT into a single alert, equivalent to a node in the graph. The second approach aims at collecting all steps for a single coherent APT into a separate subgraph representing the incident. Identifying the difference between the two approaches and having them work together in the same architecture is an important contribution of this work.

% \input{intro_old}

% The rest of the paper is organized as follows: in Section \ref{sec:model} we explain the system model of the attack graph. In Section \ref{sec:problem} we provide the background for the graph partitioning problem and our extended formulation. In Section \ref{sec:results} we show the results from public and private datasets. The paper is concluded in Section \ref{sec:conclusion}.

\section{System Description}
\label{sec:model}

\begin{figure*}[!t]
\centering
\includegraphics[width=0.95\textwidth,height=0.35\textheight]{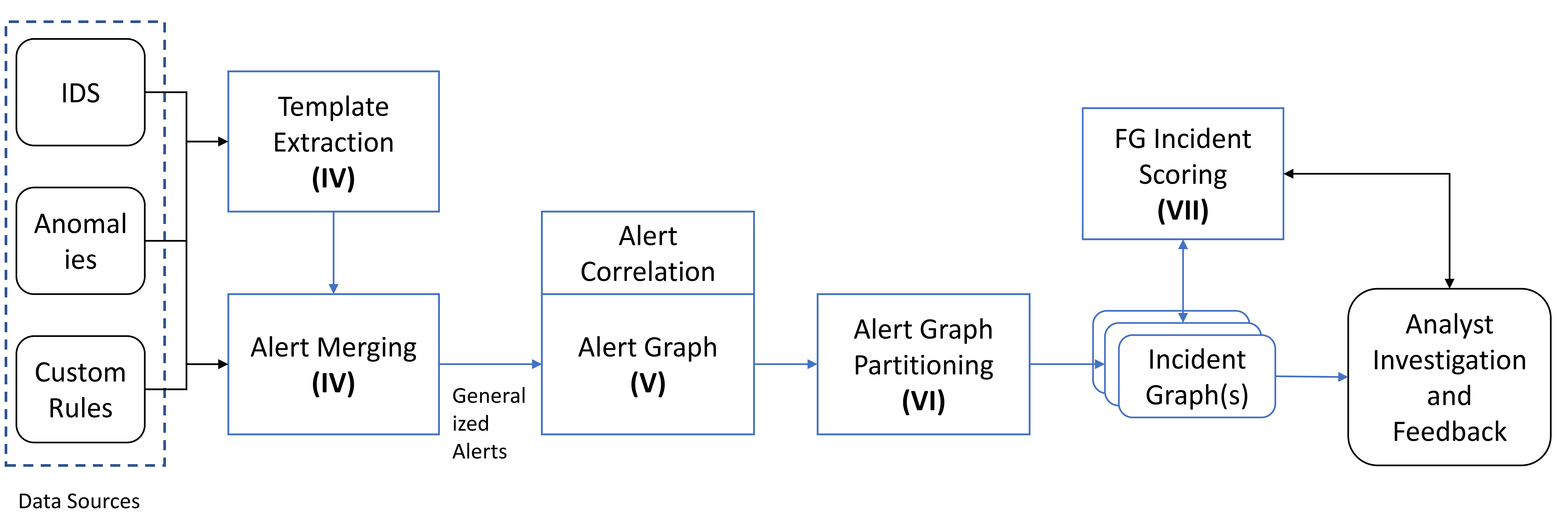}
\caption{Incident Extraction Architecture, with corresponding section labels for each step. The architectures takes as input alerts coming from a variety of sources such as IDS and custom rules. These alerts are first analyzed for common patterns and matching  alerts are merged together into generalized alerts. An alert graph is built from these generalized alerts and then partitioned into independent incident subgraphs. Resulting incident graphs are scored through their corresponding factor graphs and presented to the analyst for final investigation.}
\label{fig:architecture}
\end{figure*}

The goal of RANK is to process tens/hundreds of thousands of alerts produced by various IDS and UEBA systems into a much smaller set of proper security incidents representing truly malicious behavior. The general architecture is shown in Fig. \ref{fig:architecture}. First, we identify the inputs and outputs of the system:
\begin{itemize}
    \item \textbf{Input}: Alerts coming from sources such as Suricata\footnote{\url{https://suricata-ids.org/}} and Snort\footnote{\url{https://www.snort.org/}}, anomaly detectors and custom user-defined rules.
    \item \textbf{Output}: A number of tactic-aware security incidents, where each incident is represented as a directed graph of related generalized alerts. Each incident also includes the set of MITRE tactics present as well as a score for each tactic based on the sequence of alerts and their individual scores. Throughout the paper, we use incident and "incident graph" interchangeably.
\end{itemize}

The mapping between those inputs and outputs is accomplished through the following four stages:

\subsection{Alert Templating and Merging}
The goal of the first module in the architecture is to reduce the number of alerts by merging together alerts representing the same attack step. A \emph{generalized alert} is the result of aggregation over that particular group of alerts. For example, a port scan alert would include alerts corresponding to all ports being scanned by a single adversary. The output of this phase is a set of generalized alerts.

\subsection{Alert Correlation and Attack Graphs}
Once similar alerts have been merged together, our next objective is to establish correlation relationships between alerts. The outcome of this step is what we refer to as an alert graph. The correlation between alerts is measured using the correlation of their attributes, the time interval in-between, and how each alert fits into the MITRE ATT\&CK chain.

\subsection{Alert Graph Partitioning}
The previous stage still provides large graphs that do not necessarily lead to easily identifiable incidents. To address this issue, the alert graph is partitioned into smaller subgraphs -- incident graphs -- each representing a distinct, potential APT attack. 

\subsection{Incident Scoring}
Finally, once an incident graph has been built, our final goal is to assign a score, or set for scores, to each incident. In particular, we assign a score for each MITRE tactic present in the incident. This is achieved through a Factor Graph (FG), a type of probabilistic graphical models \cite{kschischang2001factor}. The FG considers the scores of all alerts in the incident, their sequence and their MITRE mapping, and produces a score for  each tactic present.

In the next sections, we provide details of each of the stages above.

\section{Alert Templating and Merging}
\label{sec:template}
 The first step in the RANK architecture aims at merging together alerts representing the same attack step. In this section, we follow the methodology and notation of Julisch \cite{julisch2003clustering}.

\subsection{Motivation}
Whether the alerts are coming from IDS signature, behavioral anomalies or custom rules, the end result is an overwhelming flood of alerts \cite{sommer2010outside}. However, it has been observed that there is a significant amount of redundancy in these alerts. Hence, it is our goal for the first step in the detection architecture to remove such redundancy and merge together all alerts triggered by the same underlying cause. For example, consider the Suricata alert \texttt{GPL RPC sadmind query with root credentials}. This alert can be triggered when an external adversary is scanning internal hosts for the Sadmind exploit. Even when only one external IP is scanning, the IDS will produce a separate alert for each internal host scanned. A much more concise but similarly informative alert would indicate this behavior occurring between the external adversary and all affected internal hosts. This process of merging together alerts produced by the same cause is called alert generalization \cite{julisch2003clustering}, and is the focus of this section as explained next.

\subsection{Alert Structure}
Mathematically, an alert, $\mathbf{a} \in \mathcal{A}$ with $\mathcal{A}$ being the set of alerts, is defined as a tuple of the Cartesian product
\begin{equation}
    \mathbf{a} = domB_1^{\mathbf{a}} \times domB_2^{\mathbf{a}} \times ... \times domB_{N_{\mathbf{a}}}^{\mathbf{a}}
\end{equation}
where $\{B_1^{\mathbf{a}}, B_2^{\mathbf{a}}, ... B_{N_{\mathbf{a}}}^{\mathbf{a}}\}$ is the set of alert attributes, $N_{\mathbf{a}}$ is the number of attributes in  alert $\mathbf{a}$ and $dom$ is the domain. The value an attribute $B^{\mathbf{a}}_j$ takes in alert $\mathbf{a}$ is denoted as $b^{\mathbf{a}}_j$. Examples of alert attributes are \texttt{sourceIP, destinationIP, sourcePort, destinationPort}.

% Each alert has a source, $s^{\mathbf{a}} \in \mathcal{S}$ and a score $p^{\mathbf{a}} \in [0,1]$. 
Each alert has a source, $s^{\mathbf{a}} \in \mathcal{S}$, with $\mathcal{S}$ being the set of all alerts' sources. The source of an alert refers to the rule or analysis generating this alert. For example, for an IDS alert the source is the rule ID, while for anomaly alerts the source is the name of UEBA algorithm. The set of alerts coming from a source $s$ is denoted as $\mathcal{A}_s$, and the set of attributes in all alerts in $\mathcal{A}_s$ is denoted as $\mathcal{B}_s$. All alerts coming from the same source have the same set of attributes.

Each alert also has a score $p^{\mathbf{a}} \in [0,1]$. An alert score is a measure of its severity. For example, a Suricata and Snort IDS rule has a severity entry, which can be translated into a numerical score. For UEBA-style alerts, the score is usually derived from the $p\text{-value}$ of the statistical test employed, as in \cite{sexton2015attack}.

Finally, each alert is also indicative of a set of MITRE tactics, $\mathcal{M}^{\mathbf{a}}$, and is affecting a set of internal assets/IPs $\mathcal{L}^{\mathbf{a}}$. The corresponding MITRE tactics for each alert are manually assigned based on the alert source.

\subsection{Attribute Hierarchy Trees}
Generalizing an attribute $B^{\mathbf{a}}_i$ refers to the concept of extending its domain $dom{B^{\mathbf{a}}_i}$ with a new set of values, $\{t_{1,i}, t_{2,i}, ... , t_{N_t,i}\}$, each representing a distinct subset of $dom{B^{\mathbf{a}}_i}$, resulting in the new domain $Dom{B^{\mathbf{a}}_i} = dom{B^{\mathbf{a}}_i} \cup \{t^{\mathbf{a}}_{1,i}, t^{\mathbf{a}}_{2,i}, ... , t^{\mathbf{a}}_{N_t,i}\}$.
To achieve this goal, for each attribute $B^{\mathbf{a}}_i$ we define an associated hierarchy tree $\mathcal{T}_{B^{\mathbf{a}}_i}$ representing the splitting of its domain. In this case, the number of new values $N_t$ is the number of non-leaf nodes in the corresponding hierarchy tree $\mathcal{T}_{B^{\mathbf{a}}_i}$.

\begin{figure}[!t]
\centering
\includegraphics[width=0.45\textwidth,height=0.25\textheight]{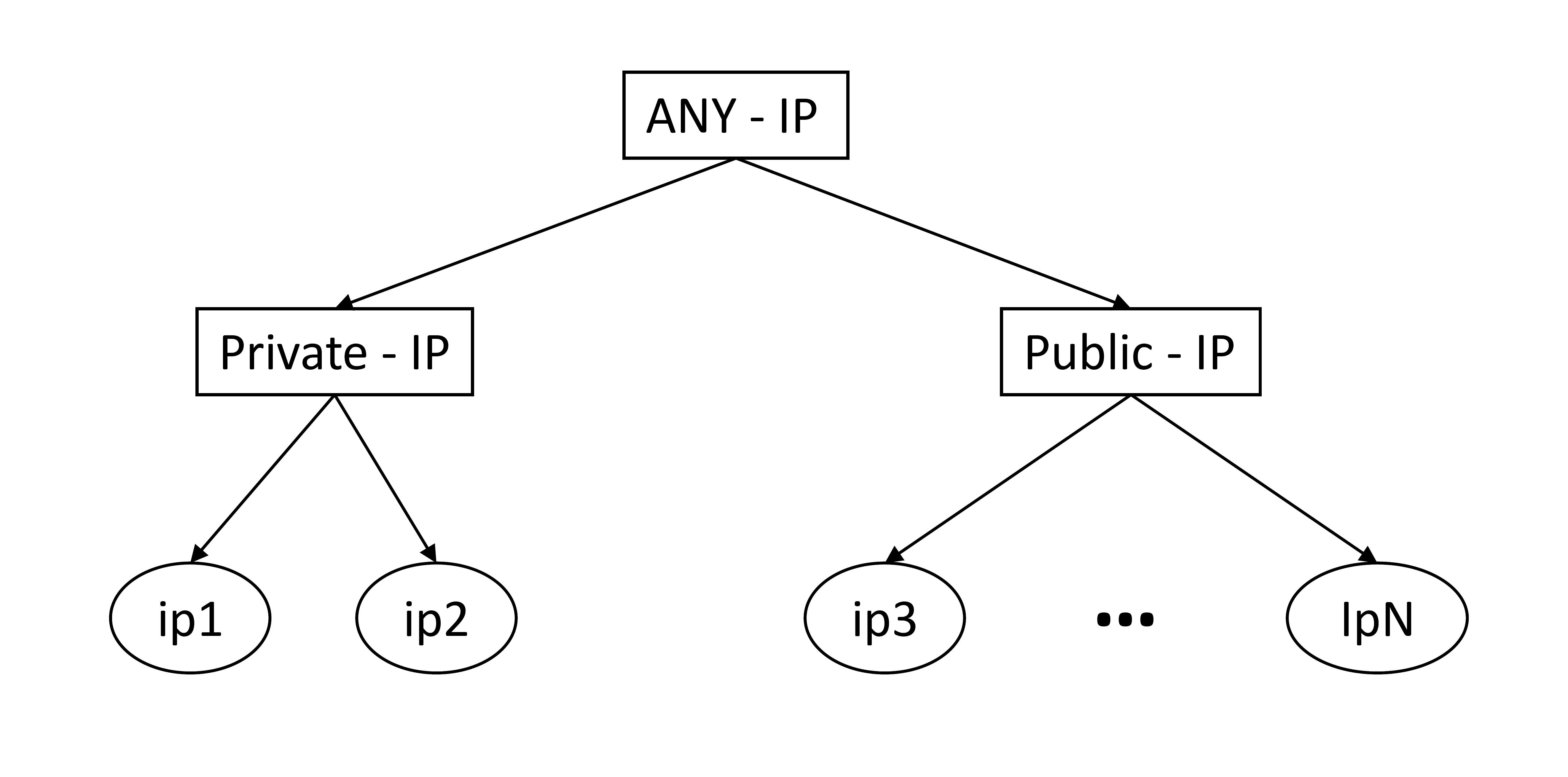}
\caption{IP Generalization Hierarchy}.
\label{fig:ip_hier}
\end{figure}

\begin{figure}[!t]
\centering
\includegraphics[width=0.45\textwidth,height=0.25\textheight]{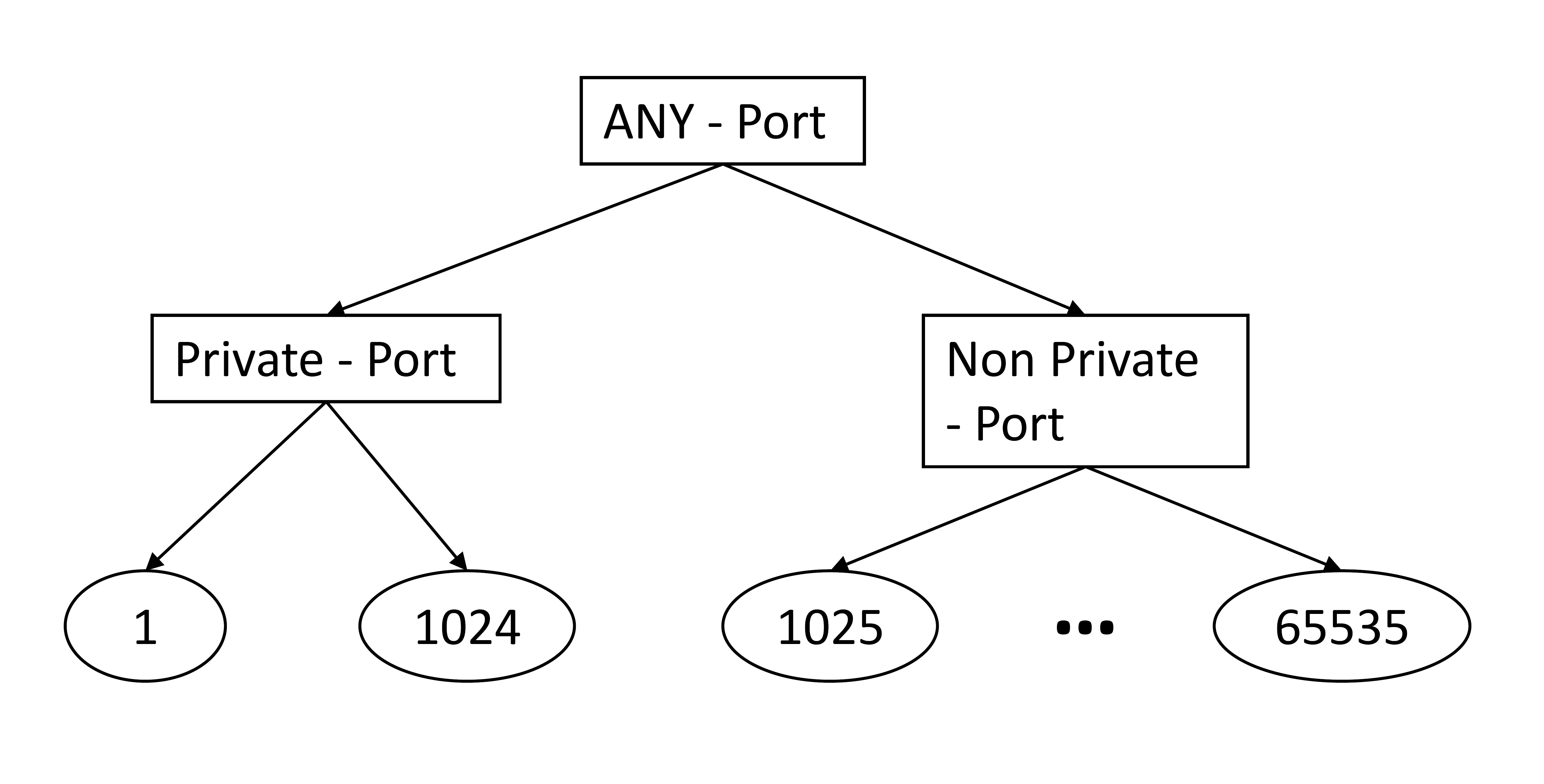}
\caption{Port Generalization Hierarchy}.
\label{fig:port_hier}
\end{figure}

An attribute $B^{\mathbf{a}}_i$ is generalized by replacing the value it takes in an alert, $b_i^{\mathbf{a}}$, with the next value in its hierarchy tree. This new value represents the subset of the attribute's domain containing the original value. Equivalently, the new attribute value becomes $b_i^{\mathbf{a}} = \textit{parent of} \; b_i^{\mathbf{a}} \; in \; \mathcal{T}_{B^{\mathbf{a}}_i}$.
We show examples of hierarchy trees for \texttt{IP} and \texttt{Port} attributes in Fig. \ref{fig:ip_hier} and \ref{fig:port_hier}, respectively. 

\subsection{Generalized Alerts}
 A generalized alert $\mathbf{v} \in \mathcal{V}$ is a tuple in the new Cartesian product
\begin{equation}
\label{eq:genalertdef}
\begin{aligned}
    \mathbf{v} = & DomB^{\mathbf{v}}_1 \times ... \times DomB^{\mathbf{v}}_{gl_s} \\ 
    & \times domB^{\mathbf{v}}_{gl_s+1} \times ... \times domB^{\mathbf{v}}_{N_{\mathbf{v}}} \\
\end{aligned}
\end{equation}
where $gl_s$ is the number of attributes to generalize per alert source and $N_{\mathbf{v}}$ is the number of attributes in the generalized alert $\mathbf{v}$. Note that generalizing an alert does not change its number of attributes. In other words, an alert is generalized when at least one of its attributes is also generalized.

It is worth emphasizing that the alert source $s^{\mathbf{a}}$ plays a crucial rule in the generalization process. First, only alerts coming from the same source can be merged together. This is done in order to keep the security context of the generalized alert meaningful. For example, if an alert is an indication of port scan for a single target machine, its generalized version will indicate a port scan for multiple target machines. The second significance of the alert source is that the parameter $gl_s$ is indexed per source not per alert. These values are manually chosen, and it is easier and more meaningful to choose the number of attributes to generalize per source. In our experiments, we have selected $gl = 2$ for IDS alerts and $gl = 1$ for UEBA alerts.

\subsection{Template Extraction and Merging Algorithm}
The high-level steps of the template extraction and alert merging are as follows:
\begin{itemize}
    \item Each alert has at least two attributes, which can include \texttt{sourceIP, destinationIP, sourcePort, destinationPort} and one or more process related attributes.
    % \item For every alert source, $s \in \mathcal{S}$, defined as the signature ID for IDS or the analysis method for UEBA, calculate $maxCount_b^s$, the number of occurrences of the most common value, for each attribute.
    \item For every alert source $s \in \mathcal{S}$ and every attribute $B_i \in \mathcal{B}_s$, calculate  the number of occurrences of the most common value $maxCount_{B_i}^s$.
    % \item Sort attributes in ascending order according to their $maxCount_b^s$.
    \item Sort attributes in ascending order according to their $maxCount_{B_i}^s$.
    \item Generalize the sorted attributes each according to its hierarchy tree until the pre-configured limit on the number of generalizable attributes. The score of the merged alert is the maximum among all the scores of its contributing individual alerts.
\end{itemize}

 In summary, the attribute to generalize at each step is the one whose most frequent value has occurred least number of times across all remaining attributes under consideration. The algorithm itself is shown in Algorithm \ref{alg:template}.


\begin{algorithm}[h]
\label{alg:template}
\SetAlgoLined
\SetKwInOut{Input}{Input}
\SetKwInOut{Output}{Output}
\SetKwProg{Init}{Initialize}{}{}
\SetKwProg{Fn}{Function}{ is}{end}

\Input{A set of alerts $\mathcal{A}$}
\Output{A set of merged and generalized alerts $\mathcal{V}$}
\Init{}{$\mathcal{V} = \mathcal{A}$}
%  \For{$\mathbf{v} \in \mathcal{V}$}{
%     $c^{\mathbf{v}} = 1$ $\;\;$// \texttt{Initialize all counts to 1}
%     } 
 \For{$s \in \mathcal{S}$ }{
    j = 0 \\
    \While{$j < gl_s$}{
  $l = SelectAttribute(\mathcal{V}_s, \mathcal{B}_s)$ \\
  \For{$\mathbf{v} \in \mathcal{V}_s$}{
  // \texttt{Generalize selected attribute}\\
    $b_l^{\mathbf{v}} = \textit{parent of} \; b_l^{\mathbf{v}} \; in \; \mathcal{T}_{l}$
    }
    \While{identical alerts $v,v' \in \mathcal{V}$ exist}{
    // \texttt{Merge matching alerts}\\
    %   $c^{\mathbf{v}} \pluseq c^{\mathbf{v}'}$\;
      $p^{\mathbf{v}} = \max (p^{\mathbf{v}}, p^{\mathbf{v}'})$ \\
      delete $v'$ from $\mathcal{V}$  % $\mathcal{V} = \mathcal{V} \ \{ v' \}$
    }
    $j \pluseq 1$
 }
}

\Fn{SelectAttribute($\mathcal{V}_s, \mathcal{B}_s)$}{
% // \texttt{The attribute to generalize next is the one whose most frequent value has occurred least number of times across all attributes under consideration} \\
% $maxCount^s_b = \max_{x \in \{ b^{\mathbf{v}} \; \forall \; \mathbf{v} \in \mathcal{V}_s\}} \text{count}(x)$ \\
%  $l = \argmin_{b \in \mathcal{B}_s} maxCount^s_b$ \\
 \For{$B_i \in \mathcal{B}_s$}{
  $maxCount_{B_i}^s = \max \left\lbrace \text{count}(x) \vert x \in \{ b_i^{\mathbf{v}} \; \vert \; \forall \, \mathbf{v} \in \mathcal{V}_s\} \right\rbrace$ 
 }
 $l = \argmin_{B_i \in \mathcal{B}_s} maxCount_{B_i}^s $ \\
 \textbf{return} $l$
}
 
 \caption{Template Extraction and Alert Merging Algorithm}
\end{algorithm}

\section{Alert Correlation and Attack Graphs}
\label{sec:corrgraph}
The previous step reduces the number of alerts to investigate by approximately two orders of magnitude. Our next step is to build an alert graph in which each node represents a generalized alert. In particular, we denote the graph $\mathcal{G} = \{ \mathcal{V}, \mathcal{E}\}$ as the alert graph, where the set of nodes $\mathcal{V}$ is the set of generalized alerts, and $\mathcal{E}$ being the set of directed edges connecting alerts together.

The main question in this section is when to add edges between the nodes, and what are the edge weights. In essence, the question is about measuring correlation between alerts. We measure correlation using two metrics: 1) correlation between the alert attributes, and 2) where each alert fits in the MITRE ATT\&CK sequence. For example, if the same external IP is targeting one or more internal hosts, with each attempted exploit triggering a different alert, then attribute correlation between said alerts would be high. Meanwhile, the MITRE ATT\&CK correlation between two alerts will increase if the second alert's tactic follows naturally from that of the first alert. For example, if the first alert represents lateral movement and the second alert represents data exfiltration, then the correlation is high. On the other hand, the correlation will be low if the first alert represents lateral movement and the second alert represents initial access. In summary, each alert is associated with one or more MITRE tactics, and the MITRE ATT\&CK correlation weight represents how likely it is for an attack to transition from the tactics associated with the first alert to the tactics associated with the second. 

% The main question in this section is when to add edges between the nodes, and what are the edge weights. In essence, the question is about measuring correlation between alerts. We measure correlation using two metrics: 1) attribute correlation between the alert, and 2) MITRE correlation. Attribute correlation should be high if alerts contain the same attribute values, while low if they are completely distinct. For example, attribute correlation would be high if the same external IP is targeting internal hosts. For MITRE correlation metric, we start by assigning one or more MITRE tactics to each alert, and then measuring how likely are the transitions between alerts based on the MITRE ATT\&CK chain. For example, if alert $v_1$ represents lateral movement and alert $v_2$ represents data exfiltration, the transition from $v_1$ to $v_2$ should be assigned higher correlation value, while a low value should be present for the case when $v_2$ is associated with initial access. 

After many experiments, we have found the following expression for the alert correlation measure, $C(\cdot, \cdot) : \mathcal{V}\times\mathcal{V} \rightarrow \mathbb{R}_{\geq0}$, to yield the best results:

\begin{equation}
\begin{aligned}
\label{eqn:corr}
& C(\mathbf{v}, \mathbf{v}') = \max_{\substack{t \in \mathcal{M}^{\mathbf{v}}\\ t' \in \mathcal{M}^{\mathbf{v}'}}}\mathbf{T}_{KC}(t,t') \; * \; \max_{\substack{ip \in \mathcal{L}^{\mathbf{v}}\\ ip' \in \mathcal{L}^{\mathbf{v}'}}} C_{\texttt{IP}}(ip,ip')\\
\end{aligned}
\end{equation}

where $C_{\texttt{IP}}(\cdot, \cdot)$ is an \texttt{IP} correlation measure, defined to be one for matching IPs and zero otherwise 
\begin{equation}
    C_{\texttt{IP}}(ip,ip') = \delta(ip,ip')
\end{equation}
where $\delta$ is the Kronecker delta function. $\mathbf{T}_{KC}(.,.)$ is a MITRE tactic transition matrix defined according to Table \ref{tab:transMat}.  We have manually assigned the transition weights and mapping, taking care to limit the number of parameters needed. The main reason behind this is the lack of properly labelled datasets containing real attacks covering many facets of the MITRE ATT\&CK matrix.

An edge $\mathbf{v} \rightarrow \mathbf{v}'$ is added to the graph $\mathcal{G}$ between the two alerts $(\mathbf{v}, \mathbf{v}')$ if their correlation measure $C(\mathbf{v}, \mathbf{v}')$ is greater than a pre-defined threshold. The threshold value has to be manually tuned and we have found a value of 0.4 to give us the best results.

\begin{table*}[hbt]
  \begin{center}
    \caption{Transition Matrix between MITRE tactics}
    \label{tab:transMat}
    \begin{tabular}{L{1.5cm} | L{0.9cm} | L{0.9cm} | L{0.9cm} | L{0.9cm} | L{0.9cm} | L{0.9cm} | L{0.9cm} | L{0.9cm} | L{0.9cm} | L{0.9cm} | L{0.9cm} | L{0.9cm}}
      \toprule % <-- Toprule here
      \textbf{TACTIC} & \textbf{Initial Access} & \textbf{Execution} & \textbf{Persistence} & \textbf{Privilege Escalation} & \textbf{Defense Evasion} & \textbf{Credential Access} & \textbf{Discovery} & \textbf{Lateral Movement} & \textbf{Collection} & \textbf{Command and Control} & \textbf{Exfiltration} & \textbf{Impact}\\
      \toprule % <-- Toprule here
      \textbf{Initial Access}  & 0.1 & 0.8 & 0.8 & 0.8 & 0.8 & 0.8 & 0.5 & 0.5 & 0.3 & 0.3 & 0.3 & 0.3 \\
      \midrule % <-- Midrule here
      \textbf{Execution} &  0.5 &  0.1 &  0.7 &  0.7 &  0.7 & 0.7 & 0.8 & 0.8 & 0.5 & 0.5 & 0.5 & 0.5  \\
      \midrule % <-- Midrule here
      \textbf{Persistence} & 0.5 & 0.7 & 0.1 & 0.7 & 0.7 & 0.7 & 0.8 & 0.8 & 0.5 & 0.5 & 0.5 & 0.5\\
      \midrule % <-- Midrule here
      \textbf{Privilege Escalation} & 0.5 & 0.7 & 0.7 & 0.1 & 0.7 & 0.7 & 0.8 & 0.8 & 0.5 & 0.5 & 0.5 & 0.5 \\
      \midrule % <-- Midrule here
      \textbf{Defense Evasion} & 0.5 & 0.7 & 0.7 & 0.7 & 0.1 & 0.7 & 0.8 & 0.8 & 0.5 & 0.5 & 0.5 & 0.5 \\
      \midrule % <-- Midrule here
      \textbf{Credential Access} & 0.3 & 0.5 & 0.5 & 0.5 & 0.5 & 0.1 & 0.8 & 0.8 & 0.5 & 0.5 & 0.5 & 0.5 \\
      \midrule % <-- Midrule here
      \textbf{Discovery} & 0.3 & 0.5 & 0.5 &   0.5 & 0.5 & 0.7 & 0.1 & 0.7  & 0.8 & 0.8 & 0.8 & 0.8 \\
      \midrule % <-- Midrule here
      \textbf{Lateral Movement} & 0.3 & 0.5 & 0.5 & 0.5 & 0.5 & 0.5 & 0.7 & 0.1 & 0.8 & 0.8 & 0.8 & 0.8 \\
      \midrule % <-- Midrule here
      \textbf{Collection} & 0.3 & 0.3 & 0.3 & 0.3 & 0.3 & 0.5 & 0.5 & 0.7 & 0.1 & 0.7 & 0.7 & 0.7 \\
      \midrule % <-- Midrule here
      \textbf{Command and Control} & 0.3 & 0.3 & 0.3 & 0.3 & 0.3 & 0.5 & 0.5 & 0.5 & 0.7 & 0.1 & 0.7 & 0.7 \\
      \midrule % <-- Midrule here
      \textbf{Exfiltration} & 0.3 & 0.3 & 0.3 & 0.3 & 0.3 & 0.5 & 0.5 & 0.5 & 0.7 & 0.7 & 0.1 & 0.7 \\
      \midrule % <-- Midrule here
      \textbf{Impact} & 0.3 & 0.3 & 0.3 & 0.3 & 0.3 & 0.5 & 0.5 & 0.5 & 0.7 & 0.7 & 0.1 & 0.1 \\
      \bottomrule % <-- Bottomrule here
    \end{tabular}
  \end{center}
\end{table*}

\section{Alert Graph Partitioning}
\label{sec:graphpartition}
With the alert graph built, the next step is to partition said graph into smaller subgraphs each representing a distinct security incident. In its general form, the graph partitioning problem is concerned with splitting the vertex set $\mathcal{V}$ of the graph $\mathcal{G}$ into a collection of non-empty subsets. The standard objective in such problems is to minimize the total weight of the edges connecting any two subsets \cite{fan2010linear}.

We utilize two approaches for graph partitioning: the ego-splitting overlapping community detection \cite{epasto2017ego}, and our proposed optimization approach \cite{Soliman2020}. The motivation behind proposing a novel approach is that community detection algorithms focus mainly on the topological properties of graphs, and fail to incorporate a security context representing the validity of the resulting sub-graphs as proper attack plans. Moreover, existing graph partitioning approaches act as a black-box and are hard to extend or improve given the security analyst feedback. 

In our previous work \cite{Soliman2020}, we have provided the first rigorous formulation of the attack partitioning problem enabling the extraction of security-aware incident graphs. Recognizing that community detection algorithms approximate the graph partitioning problem \cite{schulz2016graph}, we leveraged the power of convex optimization to incorporate security context into the problem. In particular, our objectives when partitioning a graph are:
\begin{itemize}
    \item To guarantee a proper incident, we would like to maximize the number of attack phases in each partition.
    \item To avoid impractical incidents, we would like to minimize the number of assets involved in a partition.
    \item To provide the analyst with an easy-to-visualize graph, we would like to have an upper limit on the number of alerts included in each partition.
    \item To balance how many incidents involve a single alert, our formulation allows having a reasonable and flexible limit on the number of partitions an alert might appear in. For example, limiting an alert to a single incident is overly optimistic while sharing it across many incidents is overly pessimistic. This is analogous to the concept of overlapping communities \cite{epasto2017ego}.
    \item Perhaps most importantly, to allow more flexibility in addressing security analysts feedback, we would like to have a more tractable formulation of the problem, unlike black-box community detection algorithms.
\end{itemize}

Given a graph of alerts, $\mathcal{{G} = (\mathcal{V}, \mathcal{E})}$, we define the decision matrix $\mathbf{X} = (x_{i,k})_{|\mathcal{V}|\times K}$ where $x_{i,k} \in \mathbb{Z}_2$ and $K$ is the number of partitions. The value $x_{i,k} = 1$ indicates vertex $i$ being part of partition $k$, and $0$ otherwise. We also refer to the column vectors of the matrix $\mathbf{X}$ as an incident or an incident vector $\mathbf{x}_k$. Although the number of partitions $K$ is assumed as input to our optimization problem, we note that the formulation can be extended to have $K$ as an optimization variable as well \cite{fan2010linear}.

In \cite{Soliman2020}, we showed that the overall graph partitioning optimization problem can be written as
\begin{equation}
\begin{aligned}
\label{opt:milpaftertrick}
& \underset{\mathbf{X},\mathbf{S},\mathbf{T},\mathbf{\alpha},\boldsymbol{\beta}}{\text{minimize}}
& & \gamma_0 \mathbf{e}^T\mathbf{S}\mathbf{e} + \gamma_1 \sum_{k=1}^{K} \alpha_k + \gamma_2  \sum_{k=1}^{K} \mathbf{1}^T \boldsymbol{\beta}_k\\
& \text{subject to}
& & \mathbf{X} \in \mathbb{Z}_2^{|\mathcal{V}|\times K}, \\
&&& x_{i,k} \in \mathcal{R} \cap \mathcal{P}, \\
&&& \mathbf{L}\mathbf{x}_k - \mathbf{t}_k - \mathbf{s}_k + C \mathbf{e} = 0, \\
&&& \mathbf{t}_k \leq 2 C (\mathbf{e} - \mathbf{x}_k), \\
&&& t_{i,k}, s_{i,k}  \geq 0, \\
&&& -\alpha_k\mathbf{e} \leq \mathbf{\hat{m}}_k \leq \alpha_k\mathbf{e}, \\
&&& -\boldsymbol{\beta}_k \leq \mathbf{l}_k \leq \boldsymbol{\beta}_k \;\; \forall i = 1,...,|\mathcal{V}|, k=1,...,K
\end{aligned}
\end{equation}
with
\begin{equation}
    \mathcal{R} = \left\{ x_{i,k} \in \mathbb{Z}_2 : \sum_{k=1}^{K} x_{i,k} \leq MaxMemb \right\}
\end{equation}
and
\begin{equation}
        \mathcal{P} = \left\{ x_{i,k} \in \mathbb{Z}_2 : \sum_{i=1}^{|\mathcal{V}|} x_{i,k} \leq MaxCard \right\}
    \end{equation}
where $\mathbb{Z}_2$ is the set of binary numbers, $MaxMemb$ is an upper limit on the alert incident membership, $MaxCard$ is the maximum size allowed for a partition, $\mathbf{L}$ is the graph Laplacian matrix, $C = 2 \max_{i=1,...,|\mathcal{V}|} \sum_{j=1}^{|\mathcal{V}|} w_{i,j}$ where $w_{i,j}$ are the entries of the edge weights matrix $\mathbf{W}$, $\mathbf{e}$ is an all-ones vector. A vector $\mathbf{\hat{m}}_k$ has $\hat{m}_{i,k} > 0$ if the tactic $i$ is missing from any of the alerts in incident $k$, and 0 otherwise. Similarly, a vector $\mathbf{l}_k$ has $l_{i,k} > 0$ if the asset $i$ is present in any of the alerts in incident $k$, and 0 otherwise. $\gamma_0, \gamma_1, \gamma_2$ are tunable weights for the different parts of the objective function, i.e., graph cut weight, tactic coverage and number of assets respectively. The optimization problem given in Eq. \eqref{opt:milpaftertrick} belongs to the class of mixed integer linear programming (MILP) which can be solved using commercial and open-source solvers such as CBC\footnote{\url{https://github.com/coin-or/Cbc}}. Solving problem \eqref{opt:milpaftertrick} gives us the optimum partition assignments $\hat{\mathbf{X}}$, where each partition, $\hat{\mathbf{x}}_k$, now represents a distinct security incident.

\section{Incident Scoring}
\label{sec:scoring}
With the security incidents now extracted from the bigger alert graph, the final step is to evaluate whether the sequence of alerts present in each incident represents a proper attack plan. For example, a sequence of \textbf{Initial Access} $\rightarrow$ \textbf{Execution} $\rightarrow$ \textbf{Lateral Movement} is more valid as an attack plan compared to \textbf{Lateral Movement} $\rightarrow$ \textbf{Execution} $\rightarrow$ \textbf{Initial Access}, since the first follows more closely the order in MITRE ATT\&CK. Even though the way we calculate correlation between alerts when building the alert graph tries to incorporate these relations, depending on the pair-wise correlation only is not enough. On one hand, pair-wise correlation does not look at the graph as a whole, on the other hand a single alert can be mapped to multiple MITRE tactics. Our solution here leverages factor graphs \cite{kschischang2001factor}, a type of probabilistic graphical model \cite{koller2009probabilistic}, to assign a score for each MITRE tactic present in the incident. The tactic scores are based on the scores of the individual alerts and their sequence in the incident graph.

\subsection{Overview of Factor Graphs}
\label{ssec:fg}
A factor graph is a bipartite graph used to factor a global function into a product of smaller local functions \cite{kschischang2001factor}. The graph has two types of nodes: variable nodes that represent the arguments of the function being calculated, and factor nodes defining the relationships between the variable nodes. There are three main problems under study in any application of FGs: structure, inference and learning \cite{koller2009probabilistic}. Structure is about defining the variable and factor nodes, as well as the overall topology of the graph. The inference step is concerned with designing the inference algorithm to calculate the marginal probabilities of the variable nodes. The choice of an inference algorithm is a trade-off between accuracy and feasibility, since in general the inference complexity in FG is NP-hard \cite{koller2009probabilistic}. Finally, the learning step is about the choice of parameters in the graph, such as the factor functions and the prior distribution of the variable nodes. FGs have achieved wide-success in a variety of applications such as decoding of linear codes and pattern recognition \cite{koller2009probabilistic}. 

% Moreover, a variety of algorithms in signal processing, digital communications and machine learning were shown to be instances of FGs. These include the Viterbi algorithm, Kalman filters and the Fast Fourier Transform (FFT) \cite{koller2009probabilistic}.

\subsection{Building a Factor Graph from an Incident Graph}
\label{ssec:buildfg}
Given an incident graph, i.e. a set of generalized alerts and their correlations, the question is how can we assign a score  to the different MITRE tactics present in the incident. These scores should be calculated based not on the individual alerts, rather, on the incident graph as a whole. This question can be seen as that of calculating a joint probability distribution for all alerts and tactics present in the incident. This joint probability distribution is the global function that the FG is calculating. There are many challenges here:
\begin{itemize}
    \item The number of alerts and tactics is different for each incident, the score calculation capability should be dynamic.
    \item The exponential growth in the number of states. For a binary state space of alerts $\mathcal{V}$ and tactics $\mathcal{M}$, with each state either active or inactive, the total number of states is $2^{|\mathcal{V}| + |\mathcal{M}|}$.
    \item With the lack of labelled datasets, how can we minimize the number of parameters in the FG.
\end{itemize}

In light of these challenges, we build the FG as follows:
\begin{itemize}
    \item Variables: every MITRE tactic present in the incident is mapped to a variable node.
    \item Factors: we have two classes of factor nodes
    \begin{itemize}
        \item Alert-Factors: we also have two types of alert-factors
        \begin{itemize}
            \item Single tactic alert: if an alert $\mathbf{v}$ is mapped to a single tactic, then the alert becomes a one-variable factor with $f(Active) = p^{\mathbf{v}}$ where $f(.)$ is the factor function, as shown in Table \ref{tab:singlefactor}. This factor node is connected to the corresponding variable node representing the tactic.
            \item Multi-tactic alert: if an alert $\mathbf{v}$ is mapped to multiple tactics, then the alert becomes a multi-variable factor with the same number of variables as the number of mapped tactics. The factor table entry for every possible combination of the variables' values is equal to the minimum between either the alert score or the maximum transition matrix entry between every possible pair of active tactics. The factor table is shown in Table \ref{tab:multifactor} for the case when the alert is mapped to two tactics, i.e., $\mathcal{M}^{\mathbf{v}} = \{ t_A, t_B\}$.
        \end{itemize}
        \item Transition-Factors: for every pair of tactics present in the incident, we define a new factor connecting the two. The factor function is equal to the transition matrix entry depending on the temporal order of the alerts leading to each tactic. This is shown in Table \ref{tab:pairfactor}, where $t_A \prec t_B$ means tactic $A$ happened before tactic $B$ and $\mathit{falseIndication}$ is a parameter indicating that both tactics are inactive in the incident. We chose the same value of $\mathit{falseIndication}$ for all pairs of tactics and set it to $0.2$.\\
    \end{itemize}
\end{itemize}

\begin{table}[hbt]
  \begin{center}
    \caption{Factor Table for single-mapped alert}
    \label{tab:singlefactor}
    \begin{tabular}{L{3cm} | L{3cm} }
      \toprule % <-- Toprule here
      \textbf{Tactic State} & \textbf{Factor Value} \\
      \toprule % <-- Toprule here
      Active & $p^{\mathbf{v}}$ \\
      \midrule % <-- Midrule here
      Inactive &  $1 - p^{\mathbf{v}}$  \\
      \bottomrule % <-- Bottomrule here
    \end{tabular}
  \end{center}
\end{table}

\begin{table}[hbt]
  \begin{center}
    \caption{Factor Table for multi-mapped alert}
    \label{tab:multifactor}
    \begin{tabular}{L{1cm} | L{1cm} | L{5.5cm} }
      \toprule % <-- Toprule here
      \textbf{Tactic A State} & \textbf{Tactic B State} & \textbf{Factor Value} \\
      \toprule % <-- Toprule here
      Active & Active & $\min(p^{\mathbf{v}}, \; \max(\mathbf{T}_{KC}(t_A,t_B), \mathbf{T}_{KC}(t_B,t_A)))$ \\
      \midrule % <-- Midrule here
      Inactive & Active &  $p^{\mathbf{v}}$  \\
      \midrule % <-- Midrule here
      Active & Inactive &  $p^{\mathbf{v}}$  \\
      \midrule % <-- Midrule here
      Inactive & Inactive &  1 - $p^{\mathbf{v}}$  \\
      \bottomrule % <-- Bottomrule here
    \end{tabular}
  \end{center}
\end{table}

\begin{table}[hbt]
  \begin{center}
    \caption{Pairwise Tactic Factor Table}
    \label{tab:pairfactor}
    \begin{tabular}{L{1.5cm} | L{1.5cm} | L{4cm} }
      \toprule % <-- Toprule here
      \textbf{Tactic A State} & \textbf{Tactic B State} & \textbf{Factor Value} \\
      \toprule % <-- Toprule here
    %   Active & Active & $v = \mathbf{T}_{KC}(t_A,t_B)$ if $t_A[\text{time}] <= t_B[\text{time}]$ else $\mathbf{T}_{KC}(t_B,t_A)$ \\
        Active & Active & \(\displaystyle \mu = \begin{cases} \mathbf{T}_{KC}(t_A,t_B), & t_A \prec t_B \\
 \mathbf{T}_{KC}(t_B,t_A), & \mbox{otherwise } \end{cases} \) \\
      \midrule % <-- Midrule here
      Inactive & Active &  $1 - \mu$  \\
      \midrule % <-- Midrule here
      Active & Inactive &  $1 - \mu$  \\
      \midrule % <-- Midrule here
      Inactive & Inactive &  $\mathit{falseIndication}$  \\
      \bottomrule % <-- Bottomrule here
    \end{tabular}
  \end{center}
\end{table}

This approach to building a FG for an incident has a few advantages:
\begin{itemize}
    \item The FG grows dynamically according to the sets of alerts and tactics present in the incident, circumventing the need for manual setting of the graph topology for each incident.
    \item The size of the factor tables is upper bounded by the number of MITRE tactics, i.e. 12. This limits the computational complexity of the inference algorithm and avoids dependence on the relatively larger number of generalized alerts.
    \item We use a minimal set of parameters. In particular, the set of parameters needed to build the FG are:
    \begin{itemize}
        \item The alert scores $\left\{ p^{\mathbf{v}} \; \vert \; \forall \; \mathbf{v} \in \mathcal{V} \right\}$.
        \item The tactics transition matrix $\mathbf{T}_{KC}$.
        \item The $\mathit{falseIndication}$ parameter.
    \end{itemize}
\end{itemize}

Once the FG is built, we can then run an inference algorithm on the graph to get the marginal probabilities for the variable nodes, i.e., tactics. For our case we have settled on the well-known \textit{sum-product} algorithm \cite{kschischang2001factor}. The final output of the FG is the set of marginal probabilities for all tactics present in the incident, $\left\{ P(t \; \text{is Active}) \; \forall \; t \in \mathcal{M}_k \right\}$ where $\mathcal{M}_k$ is the set of tactics in incident $k$. The incident subgraph as well as the scores of both alerts and tactics are finally presented to the security analyst for investigation.

\section{Experimental Results}
\label{sec:results}

The lack of good APT datasets, as well as the well-documented challenges of applying algorithmic and machine-learning approaches in cybersecurity \cite{sommer2010outside}, makes evaluating any work particularly challenging.  For evaluation, we use two main datasets, first is the 2000 DARPA intrusion detection dataset \cite{tavallaee2009detailed},  and the figures included in this paper are for its first scenario, with similar observations holding for the second scenario. The second dataset is a private one from a medium enterprise network. This network has around 50 employees using  primarily windows machines, as well as a few internal servers. Throughout our experiments, we used Suricata with its open rule-set  as our IDS.

\subsection{Templates}
\begin{figure}[!h]
\centerline{\includegraphics[width=0.4\textwidth,height=0.35\textheight, trim={3cm 3cm 4cm 3cm},clip]{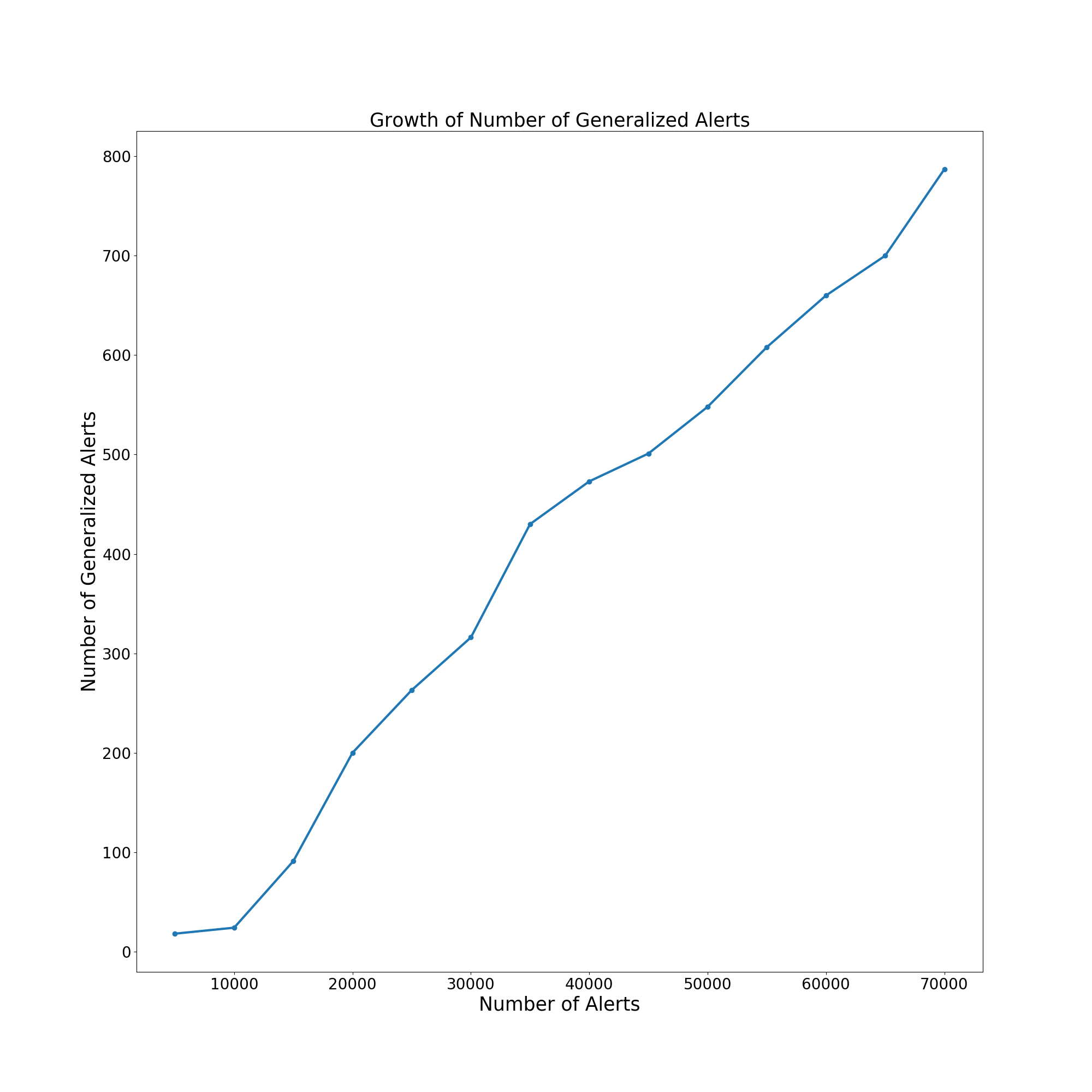}}
\caption{Reduction in number of alerts through Alert Templating and Merging}
\label{fig:templates_size}
\end{figure}

First, we study the effect of alert templating and merging on reducing the number of alerts for the later steps. In Fig. \ref{fig:templates_size}, we show the number of generalized alerts versus the number of individual alerts from the enterprise network we study. We can see a pattern where the number of generalized alerts is almost two orders of magnitude smaller than the number of individual alerts. This is quite useful by itself, as the merged alerts still keep track of all the attributes of individual ones. Meanwhile, the reduced number of alerts greatly simplifies visualization, as well as the next steps in the architecture for graph building, partitioning and scoring.

In Table \ref{tab:templates}, we show three examples of templates extracted from the DARPA dataset. The first column is the Suricata rule signature. The attributes count column means that the most common \texttt{sourcePort} value was seen only 5 times, while the most common \texttt{sourceIP} value was seen 70 times for the same data. This is expected for such an alert as it is likely that a single external adversary, \texttt{sourceIP}, is trying to gain root access in any of the enterprise hosts it can find. Hence, the template is chosen for a single \texttt{sourceIP} and single \texttt{destinationPort}, while the \texttt{destinationIP} and \texttt{sourcePort} are up-levelled according to their hierarchy tree, as seen in the second column.  

% Article top matter
\begin{table*}
\centering
\caption{Alert Templates. Number of attributes that are generalized is $gl=2$.}
\label{tab:templates}
\begin{tabular}{L{1.7cm} L{3.5cm} L{2.3cm} cccc}
\toprule
\textbf{Signature} & \textbf{Generalized Form}  & \textbf{Attributes Count}  & \multicolumn{4}{c}{\textbf{Alert Examples}} \\
\cmidrule{4-7}
 & & & \texttt{dstIP} & \texttt{dstPort} & \texttt{srcIP} & \texttt{srcPort}  \\
\midrule
\multirow{5}{\hsize}{GPL RPC sadmind query with root credentials} & \multirow{5}{\hsize}{\{\texttt{dstIP}: private-IP, \texttt{dstPort}: <port number>, \texttt{srcIP}: <IP address>, \texttt{srcPort}: Non private - Port\}} & \multirow{4}{\hsize}{(\texttt{srcPort}, 5), (\texttt{dstIP}, 30), (\texttt{dstPort}, 50), (\texttt{srcIP}, 70)} & 172.16.115.20 & 32773 & 202.77.162.213 & 60251 \\ 
    &       &  & 172.16.112.10 & 32774 & 202.77.162.213 & 60542 \\
    &       &  & 172.16.112.50 & 32773 & 202.77.162.213 & 60569 \\
    &    & & & & & \\
    &    & & & & & \\
    
\midrule
\multirow{4}{\hsize}{GPL TELNET Bad Login} & \multirow{4}{\hsize}{\{\texttt{dstIP}: <IP address>, \texttt{dstPort}: Non private - port, \texttt{srcIP}: private-IP, \texttt{srcPort}: <Port number>\}} & \multirow{4}{\hsize}{(\texttt{dstPort}, 5), (\texttt{srcIP}, 10), (\texttt{dstIP}, 20), (\texttt{srcPort}, 30)} & 195.115.218.108 & 43886 & 172.16.113.50 & 23 \\ 
    &       &  & 202.77.162.213 & 46956 & 172.16.115.20 & 23 \\
    &       &  & 202.77.162.213 & 46986 & 172.16.112.10 & 23 \\
    &    & & & & & \\

\midrule
\multirow{5}{\hsize}{GPL RPC portmap sadmind request UDP} & \multirow{5}{\hsize}{\{\texttt{dstIP}: private-IP, \texttt{dstPort}: <Port number>, \texttt{srcIP}: <IP address>, \texttt{srcPort}: Non private - Port\}} & \multirow{4}{\hsize}{(\texttt{dstIP}, 60), (\texttt{srcPort}, 60), (\texttt{srcIP}, 450), (\texttt{dstPort}, 450)} & 172.16.115.20 & 111 & 202.77.162.213 & 54790 \\ 
    &       &  & 172.16.115.87 & 111 & 202.77.162.213 & 54793 \\
    &       &  & 172.16.112.10 & 111 & 202.77.162.213 & 60540 \\
    &    & & & & & \\
    &    & & & & & \\

\bottomrule
\end{tabular}
\end{table*}

% \subsection{Alert Correlation}
% show effect of mitre mappings
% Skip probably

\subsection{Graph partitioning}
Here we study the performance of the ego-splitting framework from \cite{epasto2017ego} and our proposed approach (Section \ref{sec:graphpartition}). The ego-splitting framework utiltizes non-overlapping algorithms underneath, for which we use $\textit{Louvain}$ and $\textit{Modularity}$ from the NetworkX package\footnote{\url{https://networkx.github.io/}}, as these two have given the best results in practice. The evaluation metric is first and foremost a manual inspection of the resulting graphs, as well as statistics on the quality of those graphs, such as the average number of alerts.

We have used $\gamma_0=1.0, \gamma_1=0.5, \gamma_2=0.5, MaxMemb=2$ and $MaxCard=20$. For the experiment with the DARPA dataset we solved the integer version of problem \eqref{opt:milpaftertrick} while a relaxed version without the binary constraint is used for the experiment with the enterprise network dataset. 

\begin{figure*}[!t]
\centering
\includegraphics[width=0.95\textwidth,height=0.35\textheight]{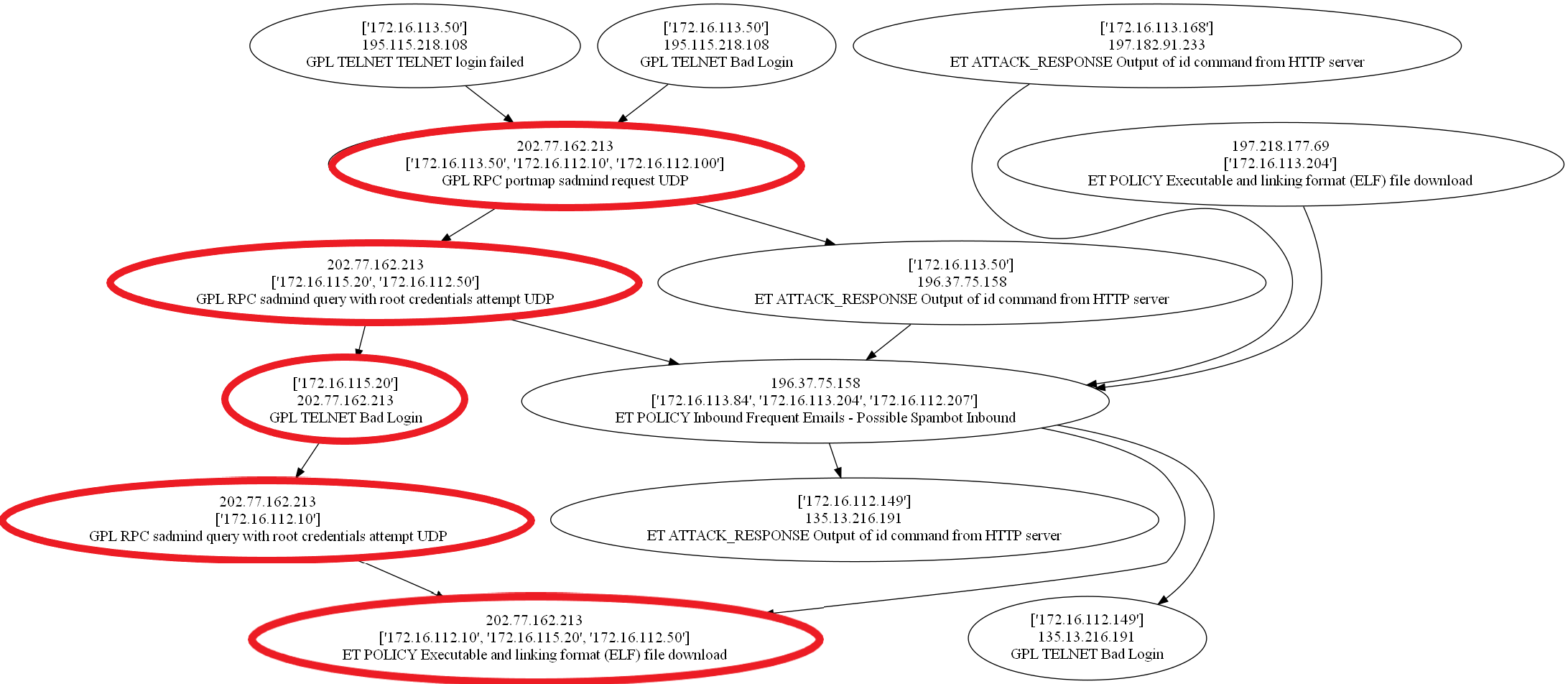}
\caption{DARPA overall alert graph prior to partitioning. The contents of each node are the source ip(s) (top), destination ip(s) (middle) and signature of the Suricata rule triggering the alert (bottom). Thick red nodes indicate the real steps of the attack. Some of the edges are omitted for visual clarity.}.
\label{fig:darpa_big}
\end{figure*}

The overall attack graph in the DARPA scenario is shown in Fig. \ref{fig:darpa_big}.  The goal of the attack is to exploit the Solaris sadmind vulnerability\footnote{\url{https://cve.circl.lu/cve/CVE-1999-0977}} on eligible hosts, install a trojan and launch a DDoS attack against a remote server. The attack progression and corresponding Suricata alerts are summarized in Table \ref{tab:darpa}.

\begin{table*}[hbt]
  \begin{center}
    \caption{DARPA Attack Progression and IDS Alerts for the case shown in Fig.~\ref{fig:darpa_big}}
    \label{tab:darpa}
    \begin{tabular}{L{5cm} | L{11cm}}
      \toprule % <-- Toprule here
      \textbf{Attack Step} & \textbf{IDS Alert(s)}\\
      \toprule % <-- Toprule here
      An IP sweep of the enterprise hosts. & The IP scan is not reported. \\
      \midrule % <-- Midrule here
      Probing the IPs from step 1 to look for sadmind hosts. & Detected by rule "GPL RPC portmap sadmind request UDP" .\\
      \midrule % <-- Midrule here
      Exploiting the Solaris Sadmind vulnerability (CVE-1999-0977). & Detected by rule "GPL RPC sadmind query with root credentials attempt UDP".\\
      \midrule % <-- Midrule here
      Installation of the DDoS trojan on exploited machines. & Detected by rule "GPL TELNET Bad Login", followed by "GPL RPC sadmind query with root credentials attempt UDP" and the actual file download in "ET POLICY Executable and linking format (ELF) file download".\\
      \midrule % <-- Midrule here
      Launching the DDoS attack. & Detected but not connected to the other alerts due to the trojans using source IP spoofing.\\
      \bottomrule % <-- Bottomrule here
    \end{tabular}
  \end{center}
\end{table*}

\begin{figure*}[!htb]
\centering
\includegraphics[width=0.95\textwidth,height=0.3\textheight]{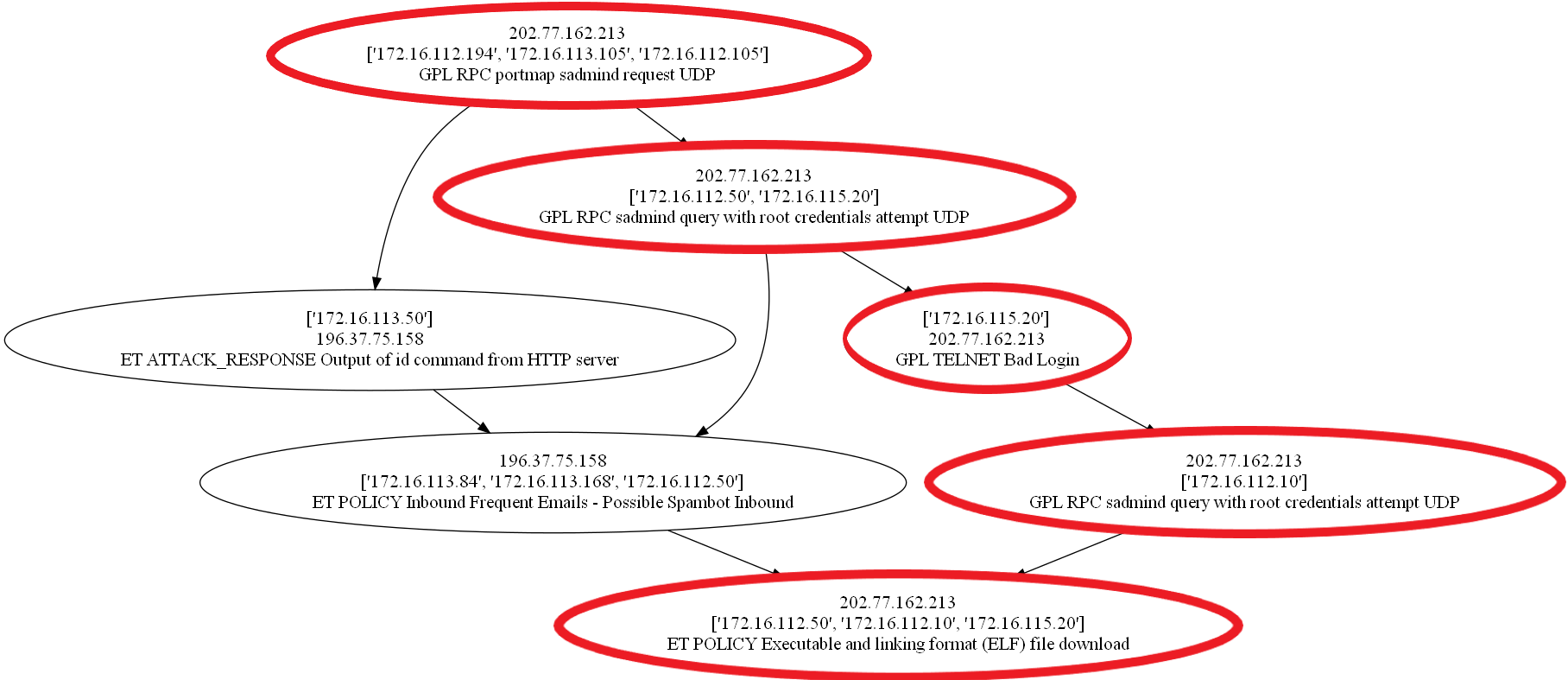}
\caption{DARPA incident graph through community detection.}
\label{fig:darpa_ego}
\end{figure*}

\begin{figure}[!htb]
\centerline{\includegraphics[width=0.5\textwidth,height=0.35\textheight]{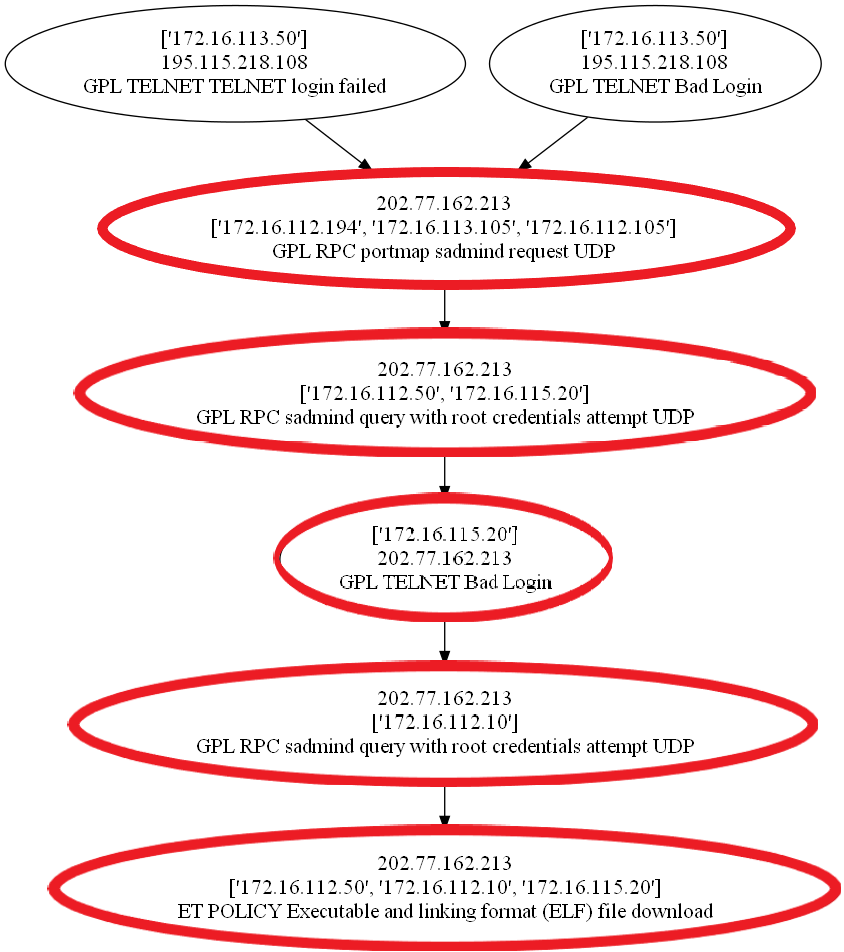}}
\caption{DARPA incident graph through graph optimization.}
\label{fig:darpa_opt}
\end{figure}

The incident extracted through our approach is shown in Fig. \ref{fig:darpa_opt}, while  the ego-splitting community detection is shown in Fig. \ref{fig:darpa_ego}.
% Both approaches have extracted all the attack steps in the incident, with our approach a single path of progression in-line with true attack behavior. 
Examining the outcomes of both approaches, we see that our method is able to extract a single path containing all attack phases which is a desirable result for the security analyst. On the other hand, the ego-splitting method contains two potential pathways from the initial to the final stage of the attack, requiring additional effort to identify the correct one.

\begin{figure}[!h]
\centerline{\includegraphics[width=0.4\textwidth,height=0.3\textheight]{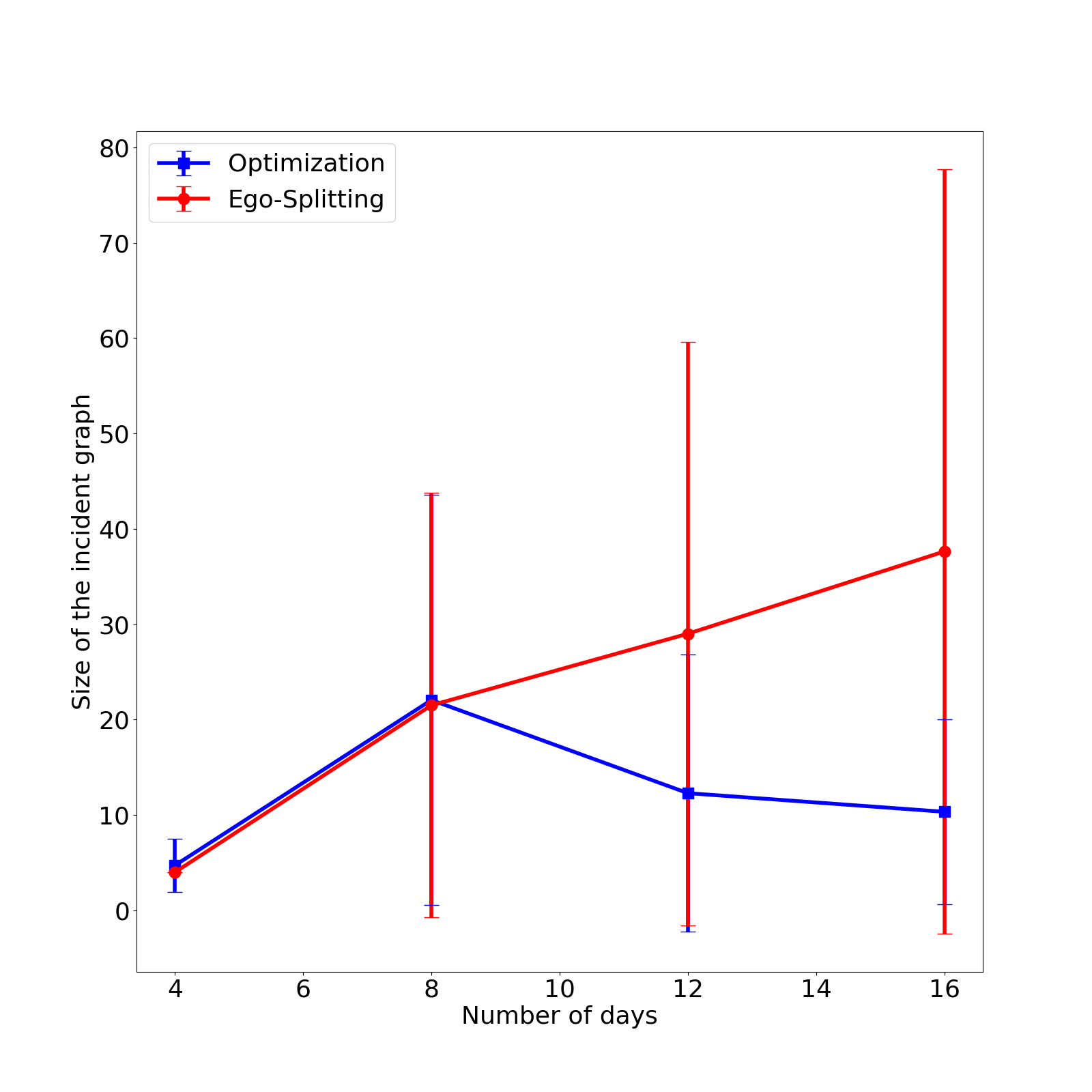}}
% \vspace{-1.0em} 
\caption{Comparison of average incident size, with vertical bars representing its standard deviation.}
\label{fig:size}
\end{figure}

For the private dataset, one interesting metric was how much the size of the incident graph grows as we lengthen the analysis period, following which the number of alerts also increases. This expansion of analysis time is necessary if we aim to discover long-term APTs. In Fig. \ref{fig:size}  we show the average incident size and its standard deviation versus the length of the analysis period. With the increase in the number of alerts, the community detection approach results in larger communities thwarting the ability of the analyst to study them, a weakness alleviated in our model. Note that we have set $MaxCard=20$ which explains the divergence of the two lines once the incident size extracted by the community detection exceeds the upper limit set in the optimization approach.

\subsection{Factor Graphs}
\begin{figure}[!htb]
\centerline{\includegraphics[width=0.5\textwidth,height=0.35\textheight]{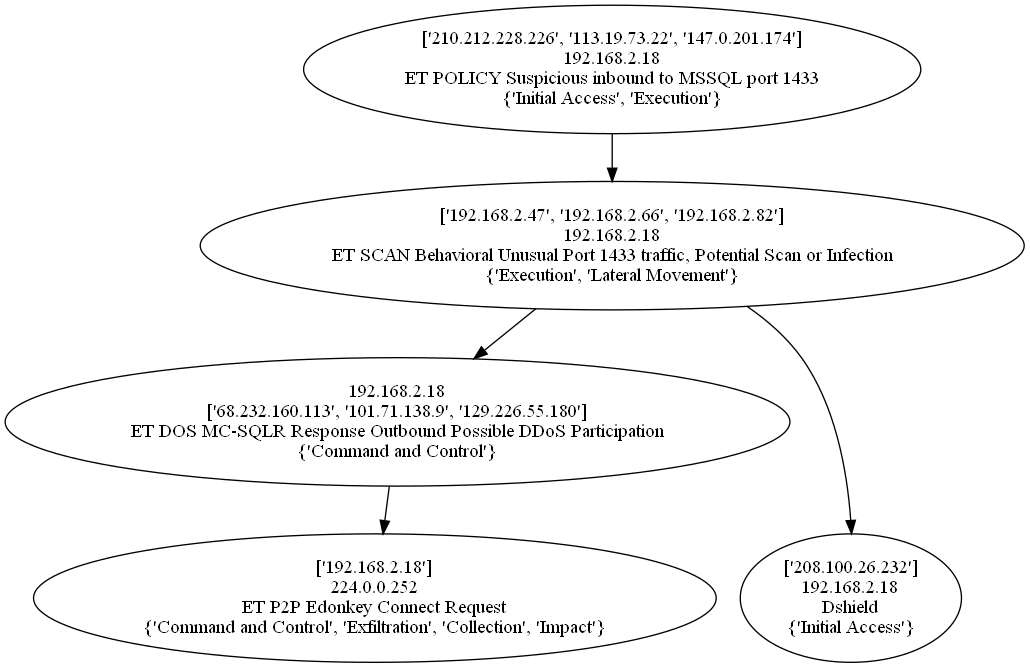}}
\caption{Factor Graph Action on an Incident. The contents of each node are the source ip(s) (top row), destination ip(s) (second row), signature of the Suricata rule triggering the alert (third row), and the MITRE tactics each alert is mapped to (bottom row).}
\label{fig:fg_incident}
\end{figure}

In Fig. \ref{fig:fg_incident} we show an example of an incident detected in our enterprise network. This incident represents exploiting an open SQL port to take part in a distributed denial-of-service (DDoS) attack. Running the belief-propagation algorithm on the FG for this incident yields the scores shown in Table \ref{tab:fg_scores}. In short, the FG rejected the \textbf{Initial Access} tactic and its alert corresponding to the branching off from the main attack path. The first two alerts are much more likely to lead to a \textbf{Command and Control} and other tactics representing the advanced progress of attack as opposed to a new attempt at accessing the network. This example shows the ability of the FG to take all alerts into consideration when scoring each individual tactic.

\begin{table}[hbt]
  \begin{center}
    \caption{FG Scores for the tactics in the incident in Fig. \ref{fig:fg_incident}}
    \label{tab:fg_scores}
    \begin{tabular}{L{3cm} | L{3cm}}
      \toprule % <-- Toprule here
      \textbf{Tactic} & \textbf{Score}\\
      \toprule % <-- Toprule here
      Initial Access & 0.0555\\
      \midrule % <-- Midrule here
      Execution & 0.9570\\
      \midrule % <-- Midrule here
      Lateral Movement & 0.9995\\
      \midrule % <-- Midrule here
      Collection & 0.9924\\
      \midrule % <-- Midrule here
      Command and Control & 0.9939\\
      \midrule % <-- Midrule here
      Exfiltration & 0.9924\\
      \midrule % <-- Midrule here
      Impact & 0.9924\\
      \bottomrule % <-- Bottomrule here
    \end{tabular}
  \end{center}
\end{table}

\subsection{Reduction}
The major benefit of the proposed architecture is reducing the amount of investigations to be done by the analyst without extensive data losses, all while automating as many parts of the process as possible and giving better representations of the data for more efficient investigation. In terms of reduction, we have achieved the following with the DARPA dataset
\begin{equation}
    7000 \; \text{Alerts} \rightarrow 44 \; \text{Generalized Alerts} \rightarrow 3 \; \text{Incidents}
\end{equation}
and for the enterprise network, for one month of data
\begin{equation}
    77000 \; \text{Alerts} \rightarrow 1100 \; \text{Generalized Alerts} \rightarrow 60 \; \text{Incidents}
\end{equation}

We observe around two orders of magnitude reduction through templating and one order of magnitude reduction through graph partitioning. The end result is a considerably more concise representation of the alerts, significantly easing the investigation burden on the analyst.

\section{Peripheral Problems}
\label{sec:periph}

In this section, we outline a few crucial areas that heavily influence the design and implementation of our architecture. These areas do not constitute a separate set of stages, instead, they tie closely to the design of the four stages discussed in Section \ref{sec:model}.

\subsection{Scalable Implementation}
In terms of implementation, the process of converting alerts into distinct incidents, each
composed of multiple generalized alerts, is divided into three main
steps:

\begin{itemize}
  \item Learning alert templates from batches of alerts.
  \item Processing alerts as they are created and merging them into generalized alerts.
  \item Constructing alert graphs from batches of generalized alerts, partitioning them into incident graphs, and scoring the incidents.
\end{itemize}
The first and second step, being independent for each alert source, lend themselves naturally to a \texttt{Map-Reduce} implementation \cite{dean2008mapreduce}. Spark\footnote{\url{https://spark.apache.org/}} is a popular choice here and is what we have used internally. The third step, focusing on graph processing, is best implemented through a message-passing framework \cite{malewicz2010pregel}. Together, these solutions enable horizontal-scaling of our architecture across a dynamic number of computing instances, and have allowed us to process data from thousands of devices in large enterprise networks.

\subsection{IP Similarity}
\label{ssec:ipsimilar}
Measuring similarity between IP addresses is an
important task in the daily operations of any enterprise network.
Applications that depend on an IP similarity measure include
measuring correlation between security alerts, building baselines
for behavioral modelling, debugging network failures and
tracking persistent attacks. For example, in our case a novel IP similarity measure $C_{\texttt{IP}}'(\cdot, \cdot)$ can be plugged into Eq. \ref{eqn:corr} to measure correlation betwen two alerts. However, IP addresses do not have a natural
similarity measure by definition. Deep Learning (DL) architectures are
a promising solution here since they are able to learn numerical
representations for IP addresses directly from data. Given these numerical representations, various
distance functions can be applied to measure similarity between the corresponding IP addresses. In a recent work \cite{burr2020detection}, we have leveraged a variant of the word2vec algorithm \cite{goldberg2014word2vec} to measure similarity between IPs learned from network log data, with promising results in incident extraction and investigation. In a follow-up work \cite{soliman2020graph}, we have extended \cite{burr2020detection} by leveraging Graph Neural Networks (GNN) for IP similarity. GNNs have the advantage  of being inductive models, i.e. they can measure similarity between new IPs not necessarily encountered during the training phase. This is a crucial feature for these models to scale to  real-world networks.

% Current works have utilized Natural Language Processing
% (NLP) techniques for learning IP embeddings. However, these
% approaches have no proper way to handle out-of-vocabulary
% (OOV) IPs not seen during training. In this paper, we propose a
% novel approach for IP embedding using an adapted graph neural
% network (GNN) architecture. This approach has the advantages
% of working on the raw data, scalability and, most importantly,
% induction, i.e. the ability to measure similarity between previously
% unseen IPs. Using data from an enterprise network, our approach
% is able to identify similarities between local DNS servers and
% root DNS servers even though some of these machines are never
% encountered during the training phase.

\subsection{Security Analyst Feedback}

The analyst can interact with the extracted incident as well as the general process of the extraction architecture in a variety of ways. These include:
\begin{itemize}
    \item An analyst can interact with the generated incident by marking either an alert or a tactic as $Inactive$, indicating a false-positive. This feedback can be handled by the FG in the form of an evidence-based query \cite{koller2009probabilistic}. An evidence-based query is an inference process on the FG with a limit on the values the conditioned variable can take.
    \item In case an analyst has marked an alert as false-positive, then a parallel process can look for similar alerts and suggest them to the analyst for further evaluation. Measuring similarity between alerts can be partially answered by measuring similarity between their IP attributes. The approaches discussed in Section \ref{ssec:ipsimilar} are of value here.
    \item The analyst may wish to revisit some of the generated templates. For example, the template generation mechanism might choose to generalize \texttt{sourceIP} and \texttt{destinationIP}. Generalizing on both IPs is not usually a desired behavior. Re-arranging the  order of attributes for templating can be easily done by the analyst afterwards.
\end{itemize}
We have only implemented the first point using evidence-based queries as will be discussed in Appendix \ref{append:evidence}, while the integration of the other two is part of our future work.

\subsection{Causality}
One important observation from discussing the incident graphs with security analysts is that an analyst expects an edge connecting two alerts to indicate a causal relationship. The question of learning causality is much more challenging than that of learning correlation \cite{pearl2009causal}. We have tried two approaches to address the problem of causality:
\begin{itemize}
    \item Protocol-based: This approach would use knowledge of the internals of different networking protocols to identify if a previous packet or flow led to the current packet or flow. As an example, a DNS query for a certain domain name together with the response containing the domain's IP address can be seen as the cause of the next HTTP traffic to that IP. The main challenge here is handling all of the possible protocol combinations expected in the network and all their variants, which is quite a challenging task.
    \item Statistical: This approach utilizes some of the established algorithms for learning a causal relations graph directly from the data \cite{spirtes1991algorithm}. However, the nature of our data and its main representation as low-level TCP/UDP flow logs hinders the applicability of algorithms such as the PC-algorithm. In particular, the layered structured of the networking protocol results in layer-3 events being all of the same type for different applications, and we cannot build a causality graph from these layer-3 flows. This approach has more success for other types of data with more context such as Syslog \cite{kobayashi2017mining}.
\end{itemize}

\section{Future Work}
\label{sec:future}

% Causality, extending factor graphs
Our planned future works centers on several lines of potential improvements:
\begin{enumerate}
    \item \textbf{Causality}: Both statistical and protocol-based approaches to measure causality between network events have proved challenging. A potential third avenue built around leveraging end-point data to determine the universally unique identifier (UUID) \cite{leach2005universally} for the software process generating each event is promising. Once the network-event to process mapping is done, all subsequent events from the same process can be seen as part of the same causal chain.
    \item \textbf{Scalablity of Graph Partitioning}: The main weakness in the optimization approach proposed for graph partitioning is its scalability, which is limited by the ability of the solvers to handle growing graph sizes. For example, we observe around 1000 generalized alerts for a three week period in a medium enterprise network. Limiting the size of each incident to 10 alerts, the resulting decision matrix has $10^6$ variables. The operational limit of open-source linear-programming solvers is between hundred thousands to one million variables, and the integer constraint in the problem quickly renders it infeasibly-large. We would like to study a more-practical heuristic that can provide a good approximation to problem \eqref{opt:milpaftertrick}.
    \item \textbf{Streaming}: In the current implementation, the graph building, partitioning and incident scoring are all run as periodic batch jobs. We plan to convert these into a streaming context, with each incoming new alert assigned to its proper incident, or dropped, and the scores of the tactics present updated accordingly.
    \item \textbf{Extending Factor Graphs}: Currently, our FG considers if a sequence of two tactics in an incident is valid from a MITRE ATT\&CK perspective. A further enhancement here is to build pairwise factors between alerts, checking if this specific sequence of alerts is valid. This is a more granular approach to build FGs, with the added granularity expected to lead to better scores. On the other hand, labelled datasets and increased computation are necessary to build and use these updated FGs.
    
\end{enumerate}

\section{Conclusion}
\label{sec:conclusion}
In this paper, we have proposed RANK, the first end-to-end architecture for detecting APTs in enterprise networks. The proposed architecture automates many parts of the data processing pipeline, such as learning alert templates, partitioning an alert graph into separate incidents, and leveraging factor graphs for automatically scoring these incidents. At the same time, the architecture is designed around preserving the analysts' crucial role in the investigation loop, by providing a highly-efficient and distilled representation of the data for easier investigating. Moreover, we have carefully integrated MITRE ATT\&CK into the various stages of our architecture, including alert correlation, alert graph partitioning and incident scoring.  Our major contribution is reducing the amount of data to be investigated by the analyst by three orders of magnitude, from 7000 alerts to 3 incidents in one example. We have also proposed novel approaches for alert correlation, alert graph partitioning, incident scoring, scalable implementation and IP similarity, and how they all tie together into the APT detection architecture.

\appendices

\section{Evidence-Based Scoring}
\label{append:evidence}

\begin{figure*}[!htb]
\centering
\includegraphics[width=0.95\textwidth,height=0.3\textheight]{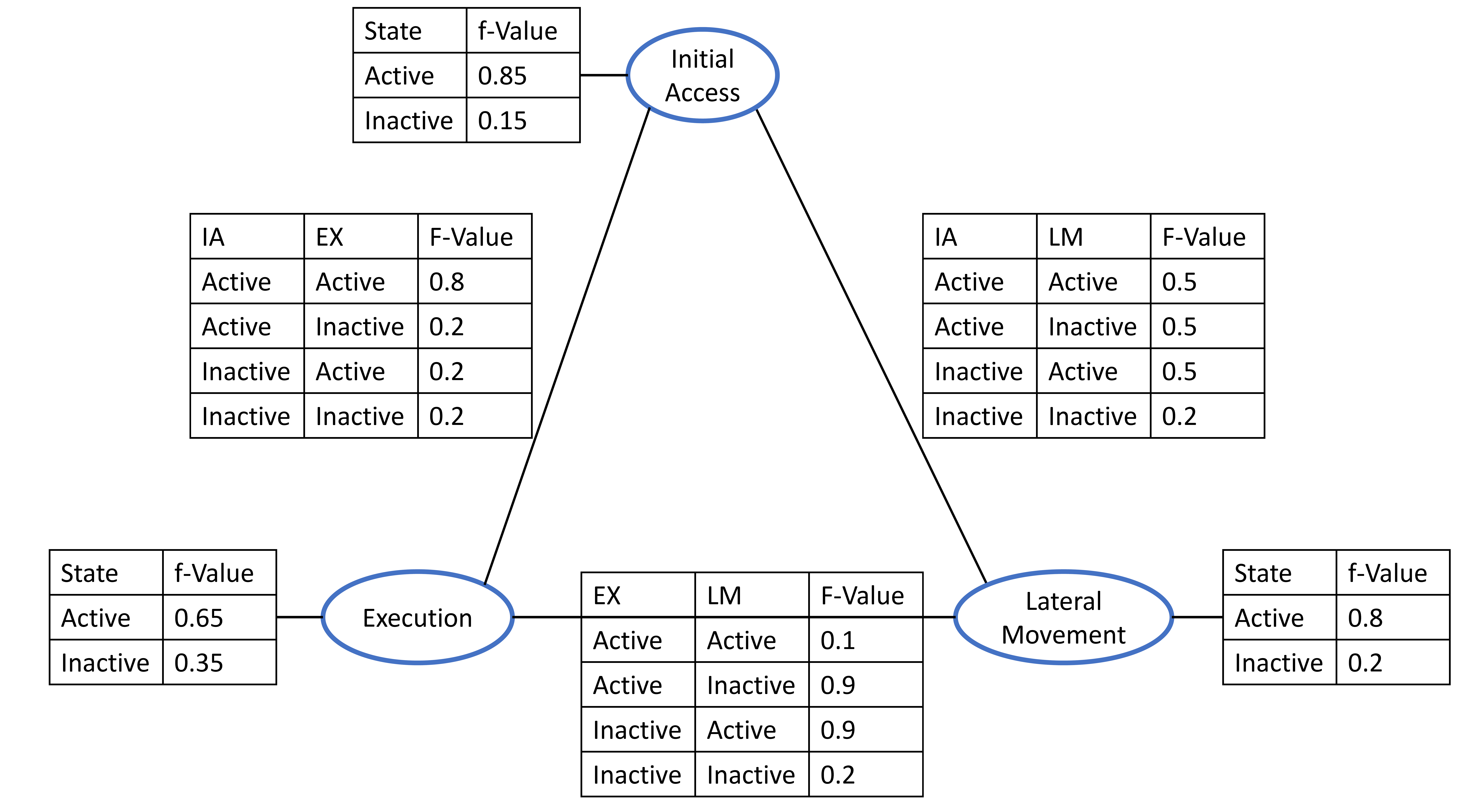}
\caption{Factor graph for a 3-tactic incident}
\label{fig:evidence}
\end{figure*}

In this section we show the effects of analyst feedback on incident scoring through evidence-based queries to the FG. In Fig. \ref{fig:evidence} we show the factor graph for a demo incident composed of three alerts. These three alerts are indicative of \textbf{Initial Access (IA)}, \textbf{Execution (EX)} and \textbf{Lateral Movement (LM)} respectively. The factor tables are calculated according to Section \ref{ssec:buildfg}. In particular, a single-variable factor represents an alert mapped to the connected tactic, while the two-variable factor represents transitions according to the matrix in Table \ref{tab:transMat}. In this incident, the temporal order of the tactics is \textbf{Initial Access (IA)} $\rightarrow$ \textbf{Lateral Movement (LM)} $\rightarrow$ \textbf{Execution (EX)}. The fact that the \textbf{Lateral Movement (LM)} happened before \textbf{Execution (EX)} is in contradiction with their order in the MITRE ATT\&CK matrix, as reflected in their factor table which penalizes both being active at the same time.

The marginal scores calculated through the \textit{sum-product} algorithm are shown in Table \ref{tab:fg_evidence_scores}. The column named \textbf{Score} is the default case before any feedback is given by the analyst. In such case, the score for \textbf{Execution (EX)} slightly increases while the score for \textbf{Lateral Movement (LM)} decreases. This is caused by the stronger reinforcement between \textbf{Initial Access (IA)} and \textbf{Execution (EX)} as compared to \textbf{Initial Access (IA)} and \textbf{Lateral Movement (LM)}. In essence, the factor graph favors an attack that progresses one step at a time over one that jumps from the first to the eighth step. The rest of the columns show the updated scores when the analyst marks a tactic as either active or inactive, i.e. true-positive or false-positive. For example, if \textbf{Execution (EX)} is marked as inactive, the score for \textbf{Lateral Movement (LM)} significantly increases due to the removal of the contradicting factor from the graph. Similar behavior is observed for \textbf{Execution (EX)} when \textbf{Lateral Movement (LM)} is the one marked as inactive instead. The stronger reinforcement between \textbf{Initial Access (IA)} and \textbf{Execution (EX)} as compared to the one between \textbf{Initial Access (IA)} and \textbf{Lateral Movement (LM)} can also be observed in the updated scores. In particular, the updated score of \textbf{Initial Access (IA)} is higher when \textbf{Lateral Movement (LM)} is the one marked as inactive compared to the case when \textbf{Execution (EX)} is marked so.

\begin{table*}[hbt]
  \begin{center}
    \caption{FG Scores for the tactics in the incident in Fig. \ref{fig:evidence} with evidence}
    \label{tab:fg_evidence_scores}
    \begin{tabular}{L{2.5cm} | L{1.5cm} | L{1.5cm} | L{1.5cm} | L{1.5cm} | L{1.5cm} | L{1.5cm} | L{1.5cm}}
      \toprule % <-- Toprule here
      \textbf{Tactic} & \textbf{Score} & \textbf{IA Active} & \textbf{IA Inactive} & \textbf{EX Active} & \textbf{EX Inactive} & \textbf{LM Active} & \textbf{LM Inactive}\\
      \toprule % <-- Toprule here
      Initial Access & 0.9374 & --- & --- & 0.9749 & 0.8540 & 0.8956 & 0.9812\\
      \midrule % <-- Midrule here
      Execution & 0.6901 & 0.7176 & 0.2772 & --- & --- & 0.4228 & 0.9695\\
      \midrule % <-- Midrule here
      Lateral Movement & 0.5111 & 0.4883 & 0.8530 & 0.3132 & 0.9519 & --- & ---\\
      \bottomrule % <-- Bottomrule here
    \end{tabular}
  \end{center}
\end{table*}

\bibliographystyle{IEEEtrannames}
\bibliography{references}

\end{document}